\documentclass[prd,reprintnumbers,twocolumn,floatfix,a4paper,nofootinbib,superscriptaddress]{revtex4-1}

\usepackage{color}
\usepackage{amsmath,amssymb,graphicx}
\usepackage{bm,lipsum}
\usepackage{times}
\usepackage{microtype}
\usepackage[utf8]{inputenc}
\usepackage{booktabs}
\usepackage{subfigure}
\usepackage[normalem]{ulem}
\usepackage[varg]{txfonts}
\usepackage{bm,graphics,graphicx,epsfig,xcolor,ulem}
\usepackage[colorlinks, pdfborder={0 0 0}]{hyperref}
\usepackage{multirow}
\definecolor{LinkColor}{rgb}{0.75, 0, 0}
\definecolor{CiteColor}{rgb}{0, 0.5, 0.5}
\definecolor{UrlColor}{rgb}{0, 0, 0.75}
\hypersetup{linkcolor=LinkColor}
\hypersetup{citecolor=CiteColor}
\hypersetup{urlcolor=UrlColor}
\allowdisplaybreaks
\newcommand{\redc}{\color{blue}}

\newcommand{\ozero}[1]{\overset{\mbox{\tiny (0)}}{#1}}

\newcommand{\otwo}[1]{\overset{\mbox{\tiny (-2)}}{#1}}
\newcommand{\othree}[1]{\overset{\mbox{\tiny (-3)}}{#1}}

\newcommand{\oone}[1]{\overset{\mbox{\tiny (-1)}}{#1}}

\newcommand{\hdpot}{{\hat \chi}}
\newcommand{\hrotpot}{{\check \chi}}

\newcommand{\Ichi}{{\mu}}

\newcommand{\lapseTB}{{N}}

\newcommand{\zhTB}{{\check h}}

\newcommand{\zhTBW}{{\gamma}}
\newcommand{\zzhTBW}{{\mathring{\zhTBW}}}

\newcommand{\zh}{{\zzhTBW }}

\newcommand{\mcN}{{\mycal N}}

\newcommand{\sectionofScri}%
{{ \,\,\,\,\mathring{\!\!\!\!\mcN}}}
\newcommand{\Bo}{\Bobo}%{{\mbox{\rm\scriptsize Bo}}}
\newcommand{\Bobo}{{}}%{\mathrm{Bo}}%{{\mbox{\rm\tiny Bo}}}

 \newcommand{\hyp}{{\mycal S}}
\newcommand{\mcL}{{\mycal L}}
\newcommand{\rbo}{r_{\Bo}}
\newcommand{\rBo}{\rbo}

\newcommand{\ringh}{{  \zzhTBW }}%

\newcommand{\bluec}{\color{blue}}

\newcommand{\eq}[1]{(\ref{#1})}

\newcommand{\eeal}[1]{\label{#1}\end{eqnarray}}

\DeclareFontFamily{OT1}{rsfs}{}
\DeclareFontShape{OT1}{rsfs}{m}{n}{ <-7> rsfs5 <7-10> rsfs7 <10-> rsfs10}{}
\DeclareMathAlphabet{\mycal}{OT1}{rsfs}{m}{n}

\definecolor{applegreen}{rgb}{0.55, 0.71, 0.0}
\definecolor{armygreen}{rgb}{0.29, 0.33, 0.13}

\definecolor{caribbeangreen}{rgb}{0.0, 0.8, 0.6}

\newcounter{mnotecount}[section]

\newcommand{\mnotex}[1]%{}
{\protect{\stepcounter{mnotecount}}$^{\mbox{\footnotesize
$%\!\!\!\!\!\!\,
\bullet$\themnotecount}}$ \marginpar{%\color{red}%
\raggedright\tiny\em
$\!\!\!\!\!\!\,\bullet$\themnotecount: #1} }

\newcommand{\bel}[1]{\begin{equation}\label{#1}}
\newcommand{\bea}{\begin{eqnarray}}
\newcommand{\bean}{\begin{eqnarray}\nonumber}
\newcommand{\beal}[1]{\begin{eqnarray}\label{#1}}
\newcommand{\eea}{\end{eqnarray}}

\newcommand{\Eq}[1]{Equation~\eq{#1}}

\newcommand{\be}{\begin{equation}}
\newcommand{\eeq}{\end{equation}}
\newcommand{\ee}{\end{equation}}
\newcommand{\beqa}{\begin{eqnarray}}
\newcommand{\eeqa}{\end{eqnarray}}
\newcommand{\beqan}{\begin{eqnarray*}}
\newcommand{\eeqan}{\end{eqnarray*}}
\newcommand{\ba}{\begin{array}}
\newcommand{\ea}{\end{array}}

\def\d{\partial}

\def\beq{\begin{eqnarray}}
\def\eeq{\end{eqnarray}}

\def\a{\alpha}
\def\b{\beta}
\def\c{\gamma}
\def\d{\delta}
\def\e{\epsilon}
\def\f{\sigma}

\def\be{\begin{equation}}
\def\ee{\end{equation}}
\def\bea{\begin{eqnarray}}
\def\eea{\end{eqnarray}}

\newcommand{\EdQuery}{\color{blue} \ptcnn{addition addressing queries of the   referees}}

\newcommand{\zspaceD}{ {\mathring D}}

\newcommand{\mcC}{\mycal C}

\newcommand{\mcH}{\mycal H}
\newcommand{\nobarg}{\blue{g}}

\newcommand{\Lie}{{\mathcal{L}}}

\newcommand{\TSxi}{{\xi}}
\newcommand{\TSr}{{r}}
\newcommand{\TSu}{{u}}
\newcommand{\TSoxi}{{\mathring \TSxi}}

\newcommand{\TSzlap}{{\Delta_{\ringh}}}%{\blue{\mathring \triangle}}

\newcommand{\TSgrn}{\color{green!70!black}}

\renewcommand{\TSgrn}[1]{#1}

\renewcommand{\bluec}{}%{\color{red} }
\renewcommand{\redc}{}%{\color{red} }
\renewcommand{\EdQuery}{}%{\color{blue} }
\renewcommand{\nobarg}{g}

\begin{document}

\title{Energy of weak gravitational waves in spacetimes  with  a positive cosmological constant
%\thanks{Preprint UWThPh-2020-12}
}
\author{P.T. Chru\'sciel} \email{piotr.chrusciel@univie.ac.at}
\affiliation{University of Vienna, Faculty of Physics, Boltzmanngasse 5, A1090 Wien, Austria}
\author{Sk Jahanur Hoque}  \email{jahanur.hoque@utf.mff.cuni.cz}
\affiliation{Institute of Theoretical Physics,
Faculty of Mathematics and Physics, Charles University,
V~Hole\v{s}ovi\v{c}k\'ach 2, 180~00 Prague 8, Czech Republic}
\author{Tomasz Smo\l{}ka} \email{tomasz.smolka@fuw.edu.pl}
\affiliation{Department of Mathematical Methods in Physics, Faculty of Physics, University of Warsaw, Pasteura 5, PL 02-903 Warszawa, Poland}

\date{\today}

\begin{abstract}
We derive a formula for the total  canonical  energy, and its flux, of weak gravitational waves on a de Sitter background.
\end{abstract}
\maketitle

In view of the recent gravitational waves detections, and of the measurement of a positive cosmological constant, there arises an urgent need of a thorough understanding of gravitational waves in spacetimes with a positive cosmological constant. An important starting point for such studies is the understanding of weak gravitational waves in this context, as these are the ones that are seen by the detectors. One of the key issues is to determine how much energy is carried away by these gravitational waves.

An attempt to provide an answer to this question using the full nonlinear Einstein equations has been made in~\cite{ChIfsits}. There a formula for the energy \emph{\`a la Trautman-Bondi} has been derived, which turned out to be ambiguous because of  a renormalised volume involved. The question then arises, whether the ambiguities  can be resolved in some cases, e.g.\ for weak field configurations.

For this,
let us consider an astrophysical system emitting gravitational waves, e.g.\ a collection of clusters of galaxies in a binary or in a localised many-body system; here ``localised'' should be understood in terms of cosmological scales. At these scales the resulting gravitational waves can be well approximated by linearised fields on a cosmological background except on a cosmologically-negligible region in an immediate neighborhood of the emitting system. And without any doubt, for such systems the field in the radiation zone is well described by the linearised theory. As a first approximation to the problem at hand it appears reasonable to consider solutions of the linearised vacuum equations throughout. The aim of this note is to derive a formula for the energy emitted in such a setting in the simplest cosmological model with a positive cosmological constant, namely de Sitter spacetime.

More precisely, we will calculate the canonical energy of weak gravitational waves on light cones in a de Sitter universe, and their flux.

Now, the worldline of an isolated system as above  is well described by a timelike geodesic in de Sitter spacetime. Such geodesics are  orbits of a Killing vector field, say $X$, which is timelike along this geodesic and tangent to it.
A first guess would be that at any moment of time along the geodesic, the total energy contained in the gravitational wave emitted equals  the integral, over the light-cone with vertex on the geodesic, of the canonical energy-momentum tensor  of the gravitational waves contracted with the vector field describing the motion of the light-cone, which in this case is the Killing vector field $X$.  However,   linearised gravitational waves are defined up to gauge transformations, and the question of a physically meaningful choice of the gauge arises. Whatever the gauge, the canonical energy-momentum tensor contracted with a Killing vector field of the de Sitter background provides a current with vanishing divergence, with the integral of this current over a three-dimensional hypersurface providing an associated charge. This is an integral over a region of infinite extent,
and will not even be finite when a random gauge is chosen. So a minimal requirement is to use gauges which allow one to obtain,  or at least identify (as will be done below),  finite integrals.
% {The gauges should be restrictive enough so that the residual gauge freedom left does not affect the total energy calculated with the method.}

In the case of vanishing cosmological constant, a good gauge for the purpose has been found by Bondi and collaborators~\cite{BBM,Sachs}. In the nonlinear theory it leads to the  Bondi energy,
as well as the associated Trautman-Bondi mass loss formula~\cite{T}.
{\bluec It provides an extremely convenient coordinate system to analyse the solutions of the characteristic constraint equations.
}
For linearised gravitational waves and $\Lambda=0$ the canonical energy, when calculated in the Bondi gauge, reproduces the Bondi energy and the Trautman-Bondi mass loss formula.
 It is therefore reasonable to expect that the use of the Bondi gauge will also provide a meaningful definition of total energy of weak gravitational waves in the presence of a positive cosmological constant. This is the approach taken here. After isolating terms which would lead to infinite energy and which have a dynamics of their own, one is thus led to our formula \eqref{8XII19.31} below for the total energy of weak gravitational waves contained in a light cone, together with the formula \eqref{8XII19.32} for the flux of energy when the cones are dragged along a timelike geodesic.

 {\EdQuery
We note that in linearised gravity the energy, as defined by the procedure below, does not depend on the gauge inside the spacetime, but only on the asymptotic behaviour of the gauge functions. So the Bondi gauge is relevant for the analysis only insofar it allows one to provide explicit formulae for   a few leading components of the   asymptotic expansion  of the metric, and give concise expressions for the integrals involved.
}

This work is related to that in~\cite{AshtekarBK,AshtekarBongaKesavanI}, where a definition of energy is used which
differs from the one used here by boundary terms.
 This difference is irrelevant when integrating over compact boundaryless surfaces. However, as made clear below, a proper inclusion of boundary terms is important to obtain the energy flux formulae that we are about to derive.
We emphasise that keeping track of these terms is crucial already in the case $\Lambda=0$, whether in the linearised case addressed here or in the full nonlinear setting.

It should be kept in mind that an approximate solution describing a gravitational wave emitted by a gravitating system has been derived in~\cite{AshtekarBongaKesavanIII}.
Unfortunately, this solution is singular near the vertex of the light cone. This is not an issue for the analysis there, since the authors of~\cite{AshtekarBongaKesavanIII} are concerned with the large-distance behaviour of the solution. However, such solutions are not suitable in our context: while the knowledge of the asymptotics of the field suffices to obtain the flux formula for energy, the solution needs to be put into the Bondi gauge to calculate our energy flux. This requires regularity everywhere, including the vertex. It would be of interest to extend the analysis in~\cite{AshtekarBongaKesavanIII} to obtain globally regular solutions.

Recall that a metric in Bondi coordinates takes the form
\begin{eqnarray}
g_{\alpha \beta}dx^{\alpha}dx^{\beta}
  &&= -\frac{V}{r}e^{2\beta} du^2-2 e^{2\beta}dudr
  \nonumber
\\
   &&+r^2\zhTBW_{AB}\Big(dx^A-U^Adu\Big)\Big(dx^B-U^Bdu\Big)
    \, ,
     \label{30XI19.100}
\end{eqnarray}
where $(x^A)\equiv(x^2,x^3) \equiv (\theta,\varphi)$,
together with the condition
\begin{equation}\label{11XII19.3}
  \det \zhTBW_{AB} = \sin^2 \theta
  \,.
\end{equation}
The de Sitter metric can  be written in this form  (cf., e.g., \cite{FischerDeSitter})
\begin{eqnarray}
{\nobarg}
&\equiv
 &{\nobarg}_{\a \b} dx^\a dx^\b
  \nonumber
\\
 &
 =
 &
  \underbrace{-(1-\Lambda r^2/3)}_{=:\epsilon\lapseTB^2}
 du^2
  -2du \, dr
 + r^2
  \underbrace{(d\theta^2+\sin^2 \theta d\phi^2)}_{=:\zzhTBW }
  \,,
   \phantom{xxx}%
%\\
%\nonumber
%& \epsilon\in\{\pm 1\}
%%\,, \quad
%% \alpha=\sqrt{{\Lambda}/{3}}
%   \,,
   \label{11XII19.1}
\end{eqnarray}
with $\epsilon\in\{\pm 1\}$ determined by the sign of $g_{uu}$.
We consider solutions, say  $h_{\mu\nu} $, of the linearised vacuum Einstein equations which, in the coordinate system of \eqref{11XII19.1}, satisfy the gauge conditions resulting from \eqref{30XI19.100}-\eqref{11XII19.3}:
\begin{equation}\label{11XII19.2}
  h_{rr}= 0 =h_{rA}
  \,,
  \quad
  \zh^{AB} h_{AB} = 0
   \,.
\end{equation}
The  equations satisfied by $h_{\mu\nu}$ can be derived from the  Lagrangian
obtained by taking the quadratic part of $\frac{\sqrt{|\det g|}}{16 \pi}\big(R- 2\Lambda \big)$
and discarding a divergence,
\begin{equation}\label{28IX18.2}
  \mcL[h]  =
  \frac 1 {32 \pi}  \sqrt{|\det g|}
  \big(
P^{ \a \b \c \d \e \f }
 \nabla_{ \a }   h_{\b \c }
 \nabla_{ \d }  h_{\e \f }
 +
  Q(h)
  \big)
  \,,
\end{equation}
where $Q$ is a quadratic polynomial in $h$,
\begin{equation}\label{6III20.6}
  Q(h)
  =
      \frac{2\Lambda}{(d-2)} \bigg[g^{\alpha\rho}g^{\beta \sigma}h_{\alpha\beta}h_{\rho\sigma}-\frac12 (g^{\alpha\beta}
      h_{\alpha\beta})^2\bigg]
%      \,,
\end{equation}
with $d=4$ here,
and
\bean
 P^{\a \b \c \d \e \f}
  & = &
  \frac 12
   \big(
    {\nobarg}^{\a \e} {\nobarg}^{\d \b} {\nobarg}^{\c \f}
    +
    {\nobarg}^{\a \e} {\nobarg}^{\f \b} {\nobarg}^{\c \d}
     -
       {\nobarg}^{\a
 \d} {\nobarg}^{\b \e} {\nobarg}^{\f \c}
  -
   {\nobarg}^{\a \b} {\nobarg}^{\c \d} {\nobarg}^{\e \f}
   \nonumber
\\
 &&
 \phantom{\frac 12 \big(}
    -
  {\nobarg}^{\b \c} {\nobarg}^{\a \e} {\nobarg}^{\f \d}
    +
     {\nobarg}^{\b \c} \nobarg^{\a \d} {\nobarg}^{\e \f}
  \big)
% \,.
\eeal{5VII18.12mine}
{\redc
(only a suitable symmetrisation of $P^{\a \b \c \d \e \f}$ is relevant, as follows from the structure of \eqref{28IX18.2}).
}
Given a Lagrangian field theory of fields $\phi^A$ with Lagrangian $\mcL$,  the
%\xout{canonical energy}
{\EdQuery Hamilton function} associated with a vector field $X$ and a hypersurface $\hyp$ is defined as the integral
\begin{eqnarray}
\mcH [\hyp, X,\phi]
 & := &
   \int_\hyp
    \big(
     \underbrace{
      \frac{\partial \mcL}{\partial \phi^A{}_\mu  } \Lie_X \phi^A - X^\mu \mcL
              }_{=: \mcH^\mu}
    \big)
    dS_\mu
    \,,
        \label{6IX18.5+}
\end{eqnarray}
where $ \phi^A{}_\mu := \partial_\mu \phi^A$.
The functional $\mcH [\hyp, X,\phi]$ is usually a starting point for calculating the Hamiltonian generating the dynamics of the theory, and equals this Hamiltonian in many cases. We shall, however, not pursue this line of thought as it involves considerations which are irrelevant for our purposes here.

Let $\mcC_{u}$ denote a level set of the coordinate $u$ as in \eqref{11XII19.1}, this is a light cone emanating from $r=0$.
 {\EdQuery
 The light cones $\mcC_{u}$ are thus obtained by moving the tip of an initial light cone $\mcC_{u_0}$ to the future  with the vector field $\partial_u$.
 In spacetimes which are asymptotically flat in null directions, for which $\Lambda=0$, the numerical value of $\mcH [\mcC_{u}, \partial_u,\phi]$ would be identified with the total energy of the field contained in $\mcC_u$. This motivates the use of the term ``energy'' when $X=\partial_u$. While the vector field $\partial_u$ is timelike near the tips of the $\mcC_{u}$'s, it is not at large distances. This last feature, unavoidable in our context, is somewhat unusual for energy for  field theory on Minkowski spacetime, but we are not on Minkowski spacetime.

  If we extend our model to one which allows matter fields concentrated near the tip of the light cone, then their contribution to the canonical energy will be the usual energy associated with a Killing vector of the background which is timelike where matter is located. This makes clear the sense in how our flux formula would include  a contribution from the usual energy of matter fields.

If attempting to carry-out our construction on a general background  we would take   $X$ to be that Jacobi field (also known as ``geodesic deviation'' vector field) along the generators of the $\mcC_{u}$'s which propagates these generators to the future without rotation at the tip; in de Sitter spacetime this Jacobi field turns out to be the restriction to $\mcC_{u}$ of the Killing vector $\partial_u$.

 Let $\mcC_{u,R}$ denote the part of the light cone $\mcC_u$ which extends from the tip to the Bondi-coordinate-radius $r=R$.
 }
Given a smooth solution $h_{\mu\nu}$ of the linearised vacuum Einstein equations
{\redc
one can find a Bondi gauge such that for small $r$ we have%
\footnote{We note that \eqref{8XI20.1} will typically not be preserved by evolution when asymptotic gauge conditions such as e.g.~\eqref{18IX18.8} are imposed, and that on nearby light cones we will then only  have $h_{AB}=O(r)$ for small $r$ in general.}
 on $\mcC_{u,R}$
\begin{equation}\label{8XI20.1}
 h_{AB} = O(r^2)
 \,.
\end{equation}
 In the  gauge \eqref{8XI20.1} one finds
}
the following formula for the canonical energy  $E_c[h,\mcC_{u,R}]$:
\begin{eqnarray}
 \lefteqn{
 E_c[h,\mcC_{u,R}]
 : =
\mcH [\mcC_{u,R}, \partial_u, h]
     }
     &&
       \nonumber
\\
 \nonumber
 &= &
  \frac{1}{64 \pi}
    \int_{\mcC_{u,R}}   {\nobarg}^{BE } {\nobarg}^{FC}
    \big(
    \partial_u h_{BC } \partial_{ r } h_{EF }
    -
     h_{BC } \partial_{ r } \partial_u h_{EF }
     \big)\, r^2 \,dr\,
      d^2\mu_{\ringh}
      % \sin(\theta) \,dr\, d\theta\,d\varphi
\\
 &&
  -
  \frac{1}{32 \pi}
      \int_{S(R) }
 P^{r (\b \c) \d (\e \f) }h_{\b\c}
 \nabla_{ \d } h_{\e \f }
   \,
      d^2\mu_{\ringh}
  \,,
  \label{30XI19.t1}
\end{eqnarray}
where $S(R)$ denotes a sphere of radius $R$ and
$$
 d^2\mu_{\ringh}
      = \sin(\theta)\, d\theta\,d\varphi
      \,.
$$
(In a  Bondi gauge where \eqref{8XI20.1} does not hold this formula cannot be used, because the fields transformed to Bondi coordinates are then too singular at the origin.)

Some readers might worry that nonlinear effects will destroy the smoothness of the light cones, making our construction inadequate for the purpose. This is not the case  in spacetimes arising from sufficiently small, fully nonlinear, perturbations on a Cauchy surface $\hyp$, for light cones emanating from points lying to the future of $\hyp$, as follows from \cite{Friedrich}.

To continue one needs to understand the asymptotic behaviour of the fields for large $r$.  In that region it is convenient to define the rescaled fields
\begin{equation}\label{13IV20.1}
 \zhTB_{\mu\nu}:= r^{-2} h_{\mu\nu }
 \,,
\end{equation}
%,
keeping in mind that the symmetric trace-free tensor field  $h_{AB}$  on a light cone is freely prescribable. It follows e.g.\ from \cite{Friedrich}
{\bluec
(compare~\cite[Proposition~2.1]{ChIfsits})
}
 that there exists a dynamically consistent class of fields $\zhTB_{AB} $ which have an asymptotic expansion of the form, for large $r$,
\begin{eqnarray}
   \zhTB_{AB}(u,r, x^C)
     &=&
   % \mathring \gamma_{AB} +
   \frac{\oone\zhTB_{AB}(u,  x^C)}{\rBo} +\frac{\otwo\zhTB_{AB}(u,  x^C)}{\rBo^2}   +\frac{\othree\zhTB_{AB}(u,  x^C)}{\rBo^3}
 \nonumber
\\
 &&      + \ldots
 \,,
 \phantom{xxx}
  \label{18IX18.8}
\label{18IX18.5}
\end{eqnarray}
%
%where $\mathring \gamma_{AB}$ is the standard metric on $S^2$ and
where the coefficients of the expansion $\oone\zhTB_{AB}$, etc., do not depend upon $\rBo$.
{\redc
Existence of a smooth conformal completion requires $\otwo\zhTB_{AB} \equiv 0$ \cite{ChIfsits},
%\ptcr{needs a comment in any case, there is a  function sigma three which is proportional to our and sigma 5; including misprints?}
and we emphasise that this is consistent with the evolution equations.
}
 The remaining $h_{\mu\nu}$'s are determined by the linearised version of the characteristic constraint equations in Bondi coordinates~\cite{MaedlerWinicour}, this proceeds as follows. %000000,CCM2}.
First, these equations give, in vacuum,  $\partial_r{h_{ru}}=0$, and %\xout{regularity at the tip of the light-cone}
{\redc
our choice of asymptotic gauge
}
leads to
$$h_{ru} \equiv 0 \,.
$$
Next we have
\begin{eqnarray}
          &&  \partial_r \left(r^4  \partial_r \zhTB_{uA} \right) =
                   \underbrace{r^2
                    \zspaceD_E
                     \left ( \zzhTBW ^{EF}\partial_r  \zhTB_{AF}\right)
                      }_{=:\psi_A}
                 \,,
                            \label{28XI19.5}
           \end{eqnarray}
where $\zspaceD_A$ is the covariant derivative
of the unit round-sphere metric $\zzhTBW_{AB}$.
Integrating in $r$ twice and using regularity of the metric at the vertex one obtains
	\begin{eqnarray}
	\zhTB_{uA}(u,r,x^A)
		& =  &
   \Ichi_A(u,x^B)
   \nonumber
\\
 &&
+
	\int_{1}^{r}\left[\frac{1}{\rho^{4}} \int_{0}^{\rho} \psi_{A}(u,s,x^A) d s\right] d \rho
  \,,
  \label{21IV20.1as}\label{6XII19.1}
	\end{eqnarray}
%
%{
%\redca{27VIII20, old}
%%
%\begin{equation}\label{6XII19.1}
%  h_{uA}(u,r,x^A) = r^2  \mu_{A} (u,x^B)   -   r^2 \int_{1}^r \psi_A(s,x^B)\left(\frac{1}{3r ^3} - \frac{1}{3s^3}\right) ds
%                 \,,
%\end{equation}
%%
%}
for  a $u$-dependent family of covector fields $\Ichi_A$  on $S^2$, to be determined shortly from the asymptotic conditions.
Equation~\eqref{21IV20.1as} leads to the asymptotic expansion
\begin{equation}\label{29XI19.2}
 \zhTB_{uA} =  \ozero \zhTB_{uA}    + \frac 12  \zspaceD^B  \oone \zhTB_{AB} \, r^{-2} +  \othree \zhTB{}_{uA}r^{-3} + \ldots
 \,,
\end{equation}
{\redc
with
\begin{equation}\label{21IV20.1}
  \othree \zhTB{}_{uA} =  - \frac 13 \lim_{r\to\infty}
  \left(
   \int_{0}^{r} \psi_{A}(s) d s
 +     \zspaceD^B  \oone \zhTB_{AB} \, r
   \right)
   \,,
\end{equation}
and we note that the limit is finite.
}

It follows from our boundary conditions and the evolution equations that the field $ \ozero h_{uA} $ is determined,
{\redc
up to a $u$-dependent family of conformal Killing vectors on the round sphere,
}
from the free characteristic data $h_{AB}$ by the equation
\begin{equation}\label{4II20.7}
  \zspaceD_A \ozero \zhTB_{uB}  +  \zspaceD_B \ozero \zhTB_{uA} -  \zspaceD^C \ozero \zhTB_{uC} \, \zh_{AB}
  = - \frac \Lambda 3 \oone \zhTB_{AB}
  \,.
\end{equation}
(This equation allows a non-zero field $\ozero\zhTB_{uA}$ even if $\Lambda=0$. In the asymptotically Minkowskian case it is consistent to assume $\ozero\zhTB_{uA}\equiv 0$,
{\redc
but it is not in general when $\Lambda >0$.)
The field $\ozero \zhTB_{uA} $ determines the field $\mu_A$ appearing in \eqref{6XII19.1}.
}
One finds  the following form of the boundary term in \eqref{30XI19.t1}:
%\ptcr{corrected 27VIII20 again, the first and third term in the first two lines in the formula in the long paper cancel out}
%
\begin{eqnarray}
 \nonumber
   &&
       -   {\frac{\Lambda R}{192  \pi} }
      \int_{S^2} \zh^{AB}\zh^{CD}\oone \zhTB_{AC}
             \oone \zhTB_{BD}
	d^2\mu_{\ringh}
     \nonumber
\\
 &&
	-
	{\frac{1}{{\TSgrn 64} \pi} }
	\int_{S^2}  \zh^{\TSgrn AB}
	\Big( \zh^{\TSgrn CD} \oone \zhTB_{AC} \partial_{u}
	\oone \zhTB_{BD}
-
	{   6  \ozero \zhTB {}_{uA} \othree \zhTB{}_{uB}}
	\Big)
\,
	d^2\mu_{\ringh}
\nonumber
\\
&&
+ o(1)
\,,
\label{30XI19.t1a}
\end{eqnarray}
where $o(1)$ denotes terms which tend to zero as $R$ tends to infinity, and where $S^2\equiv S(1)$ is the unit sphere.

The expression in \eqref{30XI19.t1a}
%diverges linearly in $R$
tends to minus infinity as   $R\to\infty$ if $\oone \zhTB_{AC}$ is not identically zero, and begs the questions whether
\begin{enumerate}
  \item   the divergence of the boundary integral is compensated by that of the volume integral  and, if not,
  \item   whether the boundary integral is needed at all in the definition of energy and, if so,
  \item  can one obtain consistent solutions by restricting oneself to a set of fields with $\oone \zhTB_{AC}\equiv 0$.
\end{enumerate}

The answer to the second  question is yes: if $\Lambda=0$, one will \emph{not} obtain  the Trautman-Bondi mass loss formula without this term.

To answer the remaining questions one needs to make use of the linearisation of the evolution equation for $g_{AB}$~\cite[Equation~(32)]{MaedlerWinicour}: Denoting by
$$
 TS[\cdot]
$$
the traceless symmetric part of a tensor, we have in vacuum
\begin{equation}\label{30XI19.12}
       r\partial_r (r   \partial_u  \zhTB_{AB})
     	 +\frac{\epsilon}{2}  \partial_r( \lapseTB^2 r^{2}  \partial_r \zhTB_{AB})
        - TS
         \big( \zspaceD_A\big( \partial_r (r^2 \zhTB_{uB}) \big)\big)
       = 0
       %8\pi        \delta T_{AB}
        \,.
\end{equation}
Integrating, we find
% \ptcr{the jump of the second power needs rechecking}
% \tscom{Checked 03II. I have put the asymptotic series from the notes (7.40-1) into the integral. Using $\delta g_{AB}=r^2 h_{AB}$ and assuming \eqref{29XI19.2}, $\otwo h_{uA}=0$, I have obtained series \eqref{30XI19.13}.}
%
\begin{eqnarray}
\lefteqn{
 \partial_u  \zhTB_{AB} (r,\cdot)
 }
 &&
 \nonumber
  \\
 & = &
  - \frac 1 r \int_{0}^r \frac{1}{s}
 \Big(
     	  \frac{\epsilon}{2}  \partial_r( \lapseTB^2 r^{2}  \partial_r \zhTB_{AB})
        - TS
         \big( \zspaceD_A\big( \partial_r (r^2 \zhTB_{uB}) \big)\big)
         \Big)(s,\cdot) ds
          \nonumber
\\
 &= &
%   \partial_u  \ozero \zhTB_{AB}(\cdot) +
    \frac {   \partial_u \oone \zhTB_{AB}(\cdot) }{r } +
    \frac {  \partial_u  \othree \zhTB_{AB}(\cdot) }{r^3}
    + o(r^{-3})
        \,,
        \label{30XI19.13}
\end{eqnarray}
where
% \ptcr{corrections; confirmed by JH on 16IV }
%
\begin{eqnarray}
  \partial_u  \oone \zhTB_{AB}(\cdot)
    & = &
   - \int_{0}^{\infty}\frac{1}{s}
 \Big(
     	   \frac{\epsilon}{2}  \partial_r[ r^2 \lapseTB^2  (\partial_r  \zhTB_{AB})]
     \nonumber
\\
 &&
         -  TS
         \big[ \zspaceD_A\big( \partial_r (r^2\zhTB_{uB}) \big)\big]
         \Big)(s,\cdot) ds
         \,.
%         \nonumber
%\\
% &&
 \label{8XII19.2}
\end{eqnarray}
{
Convergence of this integral follows from the   identity \eqref{4II20.7}.
}

\Eq{8XII19.2} shows that there is no reason for $ \partial_u  \oone \zhTB_{AB} $ to vanish in general. Hence the field $  \oone \zhTB_{AB} $ will be nonzero at later times
{\redc
even if it was on some initial light cone
}  unless the initial data are very special. Therefore the answer to question 3.\ is clearly negative.

{\EdQuery
We note that one of the issues that arises in the asymptotically flat case is the contamination of the outgoing radiation field by incoming radiation. In our construction, both when $\Lambda=0$ or $\Lambda\ne 0$, the ``no incoming radiation condition'' can be implemented by, e.g., requiring the free data on some initial light cone to be compactly supported. Formula \eqref{8XII19.2} shows that even such a strong assumption will not lead to a faster fall-off of the fields at later times in general.

Returning to our main argument,
}
 using \eqref{30XI19.13}  one finds a finite  volume contribution to the canonical energy. So the volume integral cannot be used to compensate the divergence of the boundary integral. This answers the first question in the  negative, and  is rather worrisome.

A way out is provided by the flux formula satisfied by the energy. Indeed, if $X$ is a Killing vector of the background metric $g$ and if the  field $h_{\mu\nu}$ satisfies the linearised Einstein equations, then the divergence of the field $\mcH^\mu$ vanishes. This leads to the following flux equation
\begin{eqnarray}
\lefteqn{
	\frac{d E_c[h,\mcC_{u,R}]}{du}
	=
	       }
\nonumber
 &&
\\
&&
 -{\frac{\Lambda R}{96  \pi} }
      \int_{S^2}  \zh^{AB}\zh^{CD} \oone \zhTB_{AC} \partial_{u} \oone \zhTB_{BD}
	\, d^2\mu_{\ringh}
\nonumber
\\
&&
	-
	{\frac{1}{32 \pi} }
	\int_{S^2}
	\zh^{AB}\Big(\zh^{CD}
	 	\partial_{u} \oone \zhTB_{AC}
		\partial_{u} \oone \zhTB_{BD}
-
	{   6 \othree \zhTB{}_{uA}} \partial_u	 \ozero \zhTB {}_{uB}
	\Big)
	\, d^2\mu_{\ringh}
\nonumber
\\
&&
+ o(1)
\,.
\label{30XI19.1-}
\end{eqnarray}
This equation shows that the divergent term in $E_c$ has a dynamics of its own, evolving separately from the remaining part of the canonical energy.
It is therefore natural to introduce a \emph{renormalised canonical energy}, say $\hat E_c[h,\mcC_{u,R }]$, by removing the divergent term in \eqref{30XI19.t1a}.
After having done this, we can pass to the limit $R\to\infty$ to obtain:
\begin{eqnarray}
 \lefteqn{
 \hat E_c[h,\mcC_{u }]
 : =
 }
 &&
 \nonumber
\\
 \nonumber
 &&
  \frac{1}{64 \pi}
    \int_{\mcC_{u }}   {\nobarg}^{BE } {\nobarg}^{FC}
    \big(
    \partial_u h_{BC } \partial_{ r } h_{EF }
    -
     h_{BC } \partial_{ r } \partial_u h_{EF }
     \big)\,r^2   \,dr \, d^2\mu_{\ringh}
\\
 &&
	-
	{\frac{1}{64 \pi} }
	\int_{S^2}
	  \zh^{AB} \Big(\zh^{CD}\oone \zhTB_{AC} \partial_{u}
	\oone \zhTB_{BD}
-
	{   6  \ozero \zhTB {}_{uA} \othree \zhTB{}_{uB}}
	\Big)
	\, d^2\mu_{\ringh}
  \,.
%\nonumber
%\\
% &&
  \label{8XII19.31}
\end{eqnarray}
This is our first main result here, and is our proposal how to calculate the total energy contained in a light cone of a weak gravitational wave on a de Sitter background.

The flux equation for the renormalised energy $\hat E_c$ coincides with the one obtained by dropping the term linear in $R$ in \eqref{30XI19.1-} and passing again to the limit $R\to \infty$:
%
% \ptcr{equation corrected 29IV}
%
\begin{eqnarray}
\lefteqn{
	\frac{d \hat E_c[h,\mcC_{u,R}]}{du}
	=
}
&&
 \nonumber
\\
&&
	-
	{\frac{1}{32 \pi} }
	\int_{S^2}
	\zh^{AB}\Big(\zh^{CD}
	 	\partial_{u} \oone \zhTB_{AC}
		\partial_{u} \oone \zhTB_{BD}
-
	{   6 \othree \zhTB{}_{uA}} \partial_u	 \ozero \zhTB {}_{uB}
	\Big)
	\, d^2\mu_{\ringh}
\,.
\nonumber
\\
&&
\phantom{xx}
\label{8XII19.32}
\end{eqnarray}
This is our key new formula.
When $\Lambda =0$   we recover the weak-field version of the usual Trautman-Bondi mass loss formula
{\redc
since, as already pointed out, $\ozero\zhTB_{uA}\equiv 0$ in the  asymptotically Minkowskian case.
}
Hence the last term in \eqref{8XII19.32} shows how the cosmological constant affects the flux of energy emitted by a gravitating astrophysical system.

Our mass loss formula
 has a $\Lambda$-dependent correction which can be both positive or negative, which is expected.
  Indeed, the de Sitter spacetime contains Cauchy hypersurfaces which are three-dimensional spheres, which implies that the total energy of gravitational waves in the full non-linear theory vanishes. Equivalently, in spatially closed universes  the kinetic energy of the waves is exactly compensated by the negative potential energy arising from the self-interaction of the gravitational field.
This implies in turn that any flux formula for energy must contain terms with indeterminate sign.

 It would be of interest to compare  \eq{8XII19.32} with the linearised-gravity version of the mass-evolution formula in \cite{SawPRD18}. We plan to leave this for future work. See also~\cite{SzabadosTod2} for further references.

In Appendix~\ref{A30VIII20.1}   we provide an analysis of asymptotic symmetries.
Our energy is invariant under those asymptotic symmetries which preserve \eqref{8XI20.1}.
There remains the important question of the invariance of our energy under the remaining asymptotic symmetries. There is a  difficulty here related to the fact that the gauge condition \eqref{8XI20.1}, which can be imposed on any given light cone and in which the formulae simplify, does not necessarily hold on nearby light cones. We plan to return to this question in the future.

Concluding, it should be admitted that we are still in a preliminary stage of understanding of the proper notion of energy in spacetimes with a positive cosmological constant. Our formula raises  further questions concerning its uniqueness  and its physical significance. This deserves further investigations.
 However, the analysis here gives a clear cut answer in a precise toy-model setting.

\bigskip

\noindent{\sc Acknowledgements:}
We heartfully thank Maciej Maliborski for checking the linearised Einstein equations. Useful discussions with P.~Aichelburg, T. Damour, G. %hanashyam
Date, J.~Jezierski,  M.~Kolanowski,   P. %avel
Krtou\v{s} and J.~Lewandowski are acknowledged.
JH is grateful to the Erwin Schr\"odinger Institute,
and University of Vienna for hospitality  during part of work on this paper.
%He thanks Ghanashyam Date and Pavel Krtou\v{s} for discussions.
His research was supported in part
by  the Czech Science Foundation Grant 19-01850S, and by
the DST Max-Planck partner group project ``Quantum Black Holes" between Chennai Mathematical Institute, Siruseri, Tamilnadu, India
and Albert Einstein Institute, Potsdam, Germany. TS  acknowledges the hospitality of the University of Vienna during part of work on this project and  financial support from the COST Action CA16104 GWverse. His work was supported by the University of Warsaw Integrated Development Programme (ZIP), co-financed by the European Social Fund within the framework of Operational Programme Knowledge Education Development 2016-2020, action 3.5.
The research of PTC  was   supported
by the Austrian Research Fund (FWF), Project  P 29517-N27 and
by the Polish National Center of Science (NCN) under grant 2016/21/B/ST1/00940. Last but not least, PTC is grateful to the Institut de Hautes \'Etudes Scientifiques, Bures-sur-Yvette, for hospitality during the final stage of work on this paper.

\appendix
\section{Asymptotic symmetries}
\label{A30VIII20.1}

An important question  is the gauge freedom left.  For this, one checks that vector fields $\mathring \zeta$ generating the gauge transformations
\begin{equation}\label{29II20.1}
  h_{\mu\nu}\mapsto h_{\mu\nu} + \nabla_\mu\mathring \zeta_\nu + \nabla_\nu \mathring \zeta_\mu
%  \,,
\end{equation}
and preserving the Bondi coordinate conditions as well as our asymptotic conditions
 take the form
\begin{eqnarray}
 \nonumber % Remove numbering (before each equation)
  \mathring \zeta &=&
  \left(
   \int
\frac{ \zspaceD_{B} \TSxi^{B}(u, x^{A})}{2} du
 + \TSoxi{}^u(x^ A)
  \right)\partial_u + \left(
  \frac{\TSzlap \TSxi^{\TSu}}{2} - \frac{\TSr \zspaceD_{B} \TSxi^{B}}{2}
  \right) \partial_r
   \\
    && + \big(\TSxi^B(u,x^A) - \frac{1}{\TSr} \zspaceD ^B
  \TSxi^{\TSu}(\TSu,x^A) \big)\partial_B
     \,,
      \label{20II20.9}
\end{eqnarray}
with an arbitrary function $\TSoxi^u(x^A)$ and
where, at each $u$, $ \TSxi^B(u, x^A)\partial_B  $ is a conformal Killing vector field of $\ringh$.
These gauge transformations  can be further constrained as follows:
The Hodge-Kodaira decomposition of one-forms on $S^2$ shows that there exist functions $\hdpot(u)$ and $\hrotpot(u)$ on $S^2$ such that
\begin{equation}\label{20II20.7a}
     \ozero \zhTB_{uB}(\TSu,\cdot)=\mathring{D}_{B}\hdpot(u)+{\varepsilon_{B}}^{C} \mathring{D}_{C}\hrotpot(u)
 \,,
\end{equation}
where ${\varepsilon_{B}}^{C}$ is the two-dimensional Levi-Civita tensor on the round unit two-sphere.
We can similarly write $\xi_B$ as
\begin{equation}
\xi_{B}(u,\cdot)=\mathring{D}_{B}\iota(u)+{\varepsilon_{B}}^{C} \mathring{D}_{C} \upsilon(u)
 \,,
\label{1VII20.t4a}
\end{equation}
where the functions $\iota(u)$ and $\upsilon(u)$ are linear combinations of $\ell=1$ spherical harmonics.
% (see Appendix~\ref{App2VII20}).
We have the following transformation law

\begin{equation}\label{20II20.7}
    { \ozero \zhTB_{uA}}(\TSu,x^A)+ \ringh_{AB}\partial_{\TSu} \TSoxi^{B} (\TSu,x^A)
	+ \epsilon\frac{\Lambda}3 \partial_{A} \TSxi^{\TSu}(\TSu, x^{A})
 \,,
\end{equation}
where $\epsilon$ is as in \eqref{11XII19.1},
which implies
\begin{eqnarray}\label{20II20.7ab}
  \hdpot(u)
     & \mapsto &
     \hdpot(u)
     + \partial_u \iota(u)
      +
\frac{\Lambda}3   \TSxi^{\TSu}(\TSu, \cdot)
 \,,
\\\label{20II20.7ac}
    \hrotpot(u)   & \mapsto &
      \hrotpot(u)
     + \partial_u \upsilon(u)
 \,.
\end{eqnarray}
%..
Let $P_1$ denote the $L^2(S^2)$-orthogonal projection on the space of $\ell=1$ spherical harmonics.
 We can arrange that $ P_1  \big(\hrotpot  \big)$ vanishes by solving the linear ODE
\begin{eqnarray}\label{20II20.7abcd} &
      \partial_u \upsilon
      =
      -
      P_1  \big(\hrotpot  \big)
 \,,
  &
\end{eqnarray}
which leaves the freedom of choosing $\upsilon(u_0)$.

Next, using \eqref{20II20.7ab}, together with
\begin{equation}\label{5XII19.1a}
	 \partial_{\TSu}  \TSxi^{\TSu}(\TSu,x^A) =
\frac{ \zspaceD_{B} \TSxi^{B}(\TSu, x^{A})}{2}
 \,,
\end{equation}
%\eqref{5XII19.1a}
we obtain
\begin{eqnarray}
  \partial_u\hdpot(u)
     & \mapsto &
    \partial_u \hdpot(u)
     + \partial^2_u \iota(u)
      +
\frac{\Lambda}3   \partial_u\TSxi^{\TSu}(\TSu, \cdot)
\nonumber
\\
      & = &
      \partial_u \hdpot(u)
     + \partial^2_u \iota(u)
    -
 \frac{\Lambda}3  \iota(u)
 \,.
  \label{20II20.7acc}
\end{eqnarray}
We can arrange that $\partial_u \big(P_1(\hdpot)\big)$ vanishes by solving the equation
% \ptcheck{11VII20 by TS}
%
\begin{eqnarray}  &
     \partial^2_u \iota
     -
\frac{\Lambda}3   \iota
 = P_1 \big( \partial_u \hdpot
     \big)
 \,.
  & \label{2VII20.10}
\end{eqnarray}
%.
Equation~\eqref{20II20.7ab} shows that $ P_1\hdpot $ will vanish if
\begin{eqnarray}  &
    \partial_u \iota(u_0)
      + P_1\big(
       \hdpot(u_0)
        +
\frac{\Lambda}3   \TSxi^{\TSu}(\TSu_0, \cdot)
     \big)= 0
 \,.
  & \label{2VII20.10asdf}
\end{eqnarray}
For example, Equation~\eqref{2VII20.10asdf} can be used to calculate 
$P_1( \TSxi^u(u_0,\cdot))$ after prescribing $\partial_u \iota (u_0)
$.

There remains thefore the freedom of choosing $\iota(u_0)$ and $\partial_u \iota (u_0)$,
with the solutions of the homogeneous equation \eqref{2VII20.10} taking the form
\begin{equation}\label{11VII20.1}
  \iota(u, \cdot) = e^{\alpha u} \iota_+(\cdot) + e^{-\alpha u} \iota_-(\cdot)
  \,,
\end{equation}
where $\iota_\pm$ are linear combinations for $\ell=1$ spherical harmonics.

Summarising, we can achieve a \emph{rigid transport} of the Bondi coordinates from one sphere to the other by requiring that the potentials $\hdpot$ and $\hrotpot$ of \eqref{1VII20.t4a} satisfy
\begin{equation}\label{2VII10.11}
  P_1(\hdpot)\equiv 0 \equiv P_1(\hrotpot)
  \,.
\end{equation}
We will refer to \eqref{2VII10.11} as the \emph{rigid transport condition}.

Once the rigid transport has been enforced, there remains the freedom of choosing
$\big(\iota(u_0),   \partial_u \iota (u_0),  \upsilon (u_0)\big)$, which is related to the freedom of rotating and boosting the initial light cone $\mcC_{u_0}$, and of choosing  $ \TSxi^u(u_0,\cdot)$, which is the equivalent of the supertranslations that arise in the case $\Lambda =0$, subject to the constraint
\begin{eqnarray}  &
    \partial_u \iota(u_0)
      + \frac{\Lambda}3 P_1\big(
  \TSxi^{\TSu}(\TSu_0, \cdot)
     \big)= 0
 \,.
  & \label{2VII20.10asdfb}
\end{eqnarray}

After imposing \eqref{2VII10.11}, the residual gauge transformations  which also preserve  the rigid transport condition \eqref{2VII10.11}  take the form \eqref{20II20.9}
with an arbitrary function $\TSoxi^u(x^A)$, and
where  $ \TSoxi^B (u,x^A)\partial_B  $ is the angular part of a Killing vector field of de Sitter spacetime as in \eqref{1VII20.t4a}, thus $\upsilon$ is a $u$-independent linear combination of $\ell=1$ spherical harmonics, the potential $\iota$ takes the form \eqref{11VII20.1}, with $\partial_u \iota(u_0)$ satisfying \eqref{2VII20.10asdfb}.

We will refer to these ``infinitesimal coordinate transformations'' as \emph{asymptotic symmetries}.

\bibliographystyle{amsplain}
\bibliography{PRD-format_CHS}
\end{document}